# Nonlinear change of on-axis pressure and intensity maxima positions and its relation with the linear focal shift effect

Yu. N. Makov[1], V.J. Sánchez-Morcillo[2], F. Camarena[2] and V. Espinosa[2]

[1] Department of Acoustics, Faculty of Physics, Moscow State University,

119899 Moscow, Russia.

[2] Departamento de Física Aplicada, Universidad Politécnica de Valencia, Carretera Nazaret-Oliva S/N, 46730 Grao de Gandia, Spain.

**ABSTRACT**

Experimental results related with the movement of the position of pressure and intensity maxima along the axis of focused acoustic beams under increasing driving voltages and its interpretation are presented. It is shown that in the nonlinear regime the points of the pressure and intensity maximum are separated and move differently along the axis, contrary to the linear regime, where these points coincide. The considered effects are particularly strong in the case of low-Fresnel-number beams.

## I. INTRODUCTION

The spatial localized structure of the ultrasonic field in the form of a beam has been, over the past seventy years, and stay now an object of increasing scientific attention and interest both in its fundamental and applied aspects [1-3]. The broader possibilities of the use of ultrasonic beams in practical work are associated with the ability of focusing, where the small and high intensity focal region plays a significant role for many purposes, most of them related to medical technologies [4, 5]. The on-axis distribution of the main acoustical variables, as the pressure and intensity, together with the determination of the position of their respective maxima (the largest local maximum of these two variables when several on-axis local maxima exist), provides many important information about a focused beam. These maxima are caused by the focusing effect, and their positions are usually identified with the real focus position. It is well known from the classical linear theory of focused beams [2, 3] that, in a strict sense, the position of these maxima do not coincide with the position of the geometrical focus of the system, and this discrepancy is defined by the relation between diffractional and focal lengths of the focused system. Concretely, in the linear regime the coincident position of the main on-axis pressure and intensity maxima is notably shifted towards the transducer from the geometrical focus (focal shift effect), and for the so-called low-Fresnel-number focused beams or transducers, characterized by a low value (several units or less) of the quotient between the diffractional (Rayleigh) length and the geometrical focal length $R$, this shift can be important. Although the focal shift effect is known and understandable in the linear regime of the beam propagation, in the nonlinear regime the knowledge about the evolution of this focal shift is minimal. Furthermore, the relation between the pressure and intensity focal shifts in nonlinear regime, being an important issue for many applications, has not been reported in the bibliography. All these problems are discussed in the present work on the basis of experimental results and its theoretical interpretation. We extend the results reported in [6] concerning the on-axis pressure focal shift, with a comparative analysis of the pressure and intensity focal shifts in the nonlinear regime.

In the second section the linear focal shift effect is described. We consider the linear effect as the initial condition for the main investigations concerning the evolution of the focal shift in the nonlinear regime. The description of the set-up and the experimental results are given in the third section. A discussion of the results and its theoretical interpretation is given in the fourth section. Finally, in the last section we present the conclusion.

## II LINEAR FOCAL SHIFT.

The difference between the position of the on-axis pressure main maximum and that of the geometrical focus is well understood in the framework of the linear theory. Diffraction, inherent to any wave beam, results in a broadening of the beam as it propagates along the axis, expands the focal waist (therefore decreasing the field maximum amplitude at the waist) and moves its position towards the transducer, i.e. to the side where diffractional beam expansion is smaller. Also in the linear regime, for a sinusoidal waveform, the on-axis position of pressure and intensity main maximum coincide. The starting point of our investigation is the linear situation with focal shift, and we consider its further nonlinear evolution.

The magnitude of the linear focal shift can be determined from the spatial distribution of the pressure $p(r,z,t) = A(r,z)e^{ikz-i\omega t}$, whose complex amplitude $A(r,z)$ is the solution of

the ordinary wave equation in the parabolic approximation. This solution is (see for example [2])

$$A(r,z) = -\frac{ik}{z}\exp\left(\frac{ik}{2z}r^2\right)\int_0^\infty \exp\left(\frac{ik}{2z}r'^2\right)J_0\left(\frac{k}{z}rr'\right)A(r',0)r'dr', \qquad (1)$$

where $z$ is the longitudinal coordinate along the beam axis, $r$ is the radial (transverse) coordinate, $k$ is the wave number and $A(r, 0)$ is determined by the initial condition.

In the simplest case of constant pressure $p_0$ along the transducer surface with radius $a$ and parabolic phase profile accounting for the focusing effect, the initial condition reads

$$A(r',0) = p_0 \exp\left(-\frac{ikr^2}{2R}\right) \quad \text{for} \quad r \leq a, \qquad (2)$$

$$A(r',0) = 0 \quad \text{for} \quad r > a.$$

The on-axis pressure distribution under this initial condition can be calculated from Eq. (1):

$$\frac{p(0,\tilde{z})}{p_0} = \left|\frac{2}{1-\tilde{z}}\sin\left(\frac{\pi N_F}{2}\frac{1-\tilde{z}}{\tilde{z}}\right)\right|, \qquad (3)$$

where the bars denote absolute value, $\tilde{z} = z/R$ is a dimensionless coordinate along the beam axis and $N_F = \frac{L_d}{\pi R}$ is the Fresnel number of the focused acoustic beam, characterizing the opposite action of the broadening effect of diffraction (related to the Rayleigh length $L_d = ka^2/2$) and focusing effects (related to the focal length $R$). Taking the derivative of Eq. (3) and equating it to zero, the locations of the points of pressure extremum can be found as the solutions of the transcendental equation

$$\frac{tg(X)}{X/\tilde{z}} = 1, \qquad (4)$$

where we have defined $X = \frac{\pi N_F}{2}\frac{1-\tilde{z}}{\tilde{z}}$.

The root of Eq. (4) for which the pressure distribution in Eq. (3) takes the largest value, corresponds to the location of the main maximum of on-axis pressure, the so-called focal point. We recall that these results are valid only in the linear regime of beam propagation, where a sinusoidal temporal wave profile propagates without distortion. In this case all the results and conclusions about the on-axis pressure distribution can be extended to the on-axis intensity $I_T(\tilde{z})$, defined as

$$I_T(\tilde{z}) = \frac{1}{T} \int_{t_0}^{t_0+T} \frac{p(t,\tilde{z})^2}{\rho_0 c_0} dt, \qquad (5)$$

where $\rho_0 c_0$ is the acoustic impedance of the medium and $T$ is a period. Clearly, in the linear regime the intensity at any point $I_T(\tilde{z})$ is proportional to the square of the maximum pressure distribution (given by Eq. (3)) and therefore the maximum on-axis pressure and intensity points are located at the same axial position.

Figure 1 shows dependence of the position of the main pressure maximum ($\tilde{z}_{\max}$) with the Fresnel number $N_F$, as obtained numerically from Eq. (4). It evidences that, for $N_F > 6$ (corresponding to high-Fresnel-number focused beams) the main pressure and intensity maxima are almost coincident with the geometrical focus, while for $N_F < 3$ (low-Fresnel-number focused beams) the difference between these points is large, existing an important displacement towards the transducer of these maxima. The prediction of a strong shift of the maximum has been demonstrated experimentally in [6], using a serial (Valpey-Fisher) focused transducer with $a$ = 1.5 cm and $R$ = 11.7 cm with resonant frequency $f$ = 1 MHz inmersed in water ($\lambda$ = 0.15 cm), for which $N_F$ = 1.28. According to the previous theoretical results, the main pressure and intensity maxima in linear regime are located at a distance $0.67R$ = 7.8 cm from the transducer surface, in good agreement with the measured value [6]. This large initial (linear) focal shift allows us to study in detail the evolution (movement) of the maximum over a large range in the nonlinear regime.

### III. NONLINEAR EVOLUTION OF THE ON-AXIS PRESURE AND INTENSITY MAXIMUM POSITIONS (EXPERIMENTAL ANALYSIS).

The experimental setup followed the classical scheme of confronted emitting transducer (with the parameters indicated above) and receiving calibrated membrane hydrophone (NTR/Onda Corp. MH2000B) in a water tank with dimensions 25x25x50 cm³. The transducer was driven by the signal provided by a programmable Agilent 33220 function generator, amplified by a broadband RF power amplifier either (depending on the power requirements) ENI 240L (40W, +30dB) or ENI 500A (500W, +60 dB), which permitted to deliver voltage amplitudes at the transducer terminals up to 750 Vpp without distortion, using an impedance matching filter.

The emitter and the hydrophone were aligned using a positioning system in order to determine the symmetry axis of the system. Once the axis was defined, the pressure temporal waveforms were measured at different positions along this axis, for increasing transducer voltages ranging from 100 to 500 Vpp. From the measured waveforms the peak pressure distributions can be readily obtained (see Fig. 2). The temporal waveforms change along the beam axis at a fixed transducer voltage (due to the dynamic nonlinear profile transformation in the wave propagation) and also at a fixed position under the increase in the transducer voltage. The recording of the temporal waveforms allows to evaluate the nonlinearity degree (for example, the condition and moment of shock front formation) and also to determine, on the basis of Eq. (5), the acoustic intensity at every point on the axis (i.e. on-axis intensity distribution) and at every level of the input transducer voltage.

The experimental results are presented in Fig. 2. Figure 2(a) shows the on-axis pressure (dashed curves) and the corresponding intensity (solid lines) distributions, as the input voltage is increased from 200 to 500 Vpp. Examples of the waveforms measured at two positions for different input values are shown in Figs. 2(b) and 2(c). The maximum values of the pressure and intensity distributions for each driving voltage are denoted with black circles and squares, respectively. Two qualitatively different regimes can be observed in the behaviour of the positions of the maxima in Fig. 2(a). When the input voltage increases from 200 to 250 Vpp, the point of maximum pressure moves forward, approaching the neighbourhood of the geometrical focus. The wave profiles in this range experience an incremental nonlinear distortion but without shock formation. With a further increase of the input voltage from 300 to 500 Vpp the position of the maximum pressure shifts to the opposite direction (i.e. towards the transducer). In this regime, shock waves develop in the waveforms at a given distance. In our experiment, the transition between both regimes occurs at 250 Vpp. The intensity distribution curves in Fig. 2(a) show that the range of movement of the maximum intensity point is narrower than that for the maximum pressure point. We recall that although the maximum intensity and pressure points coincide on the axis (are located at the same coordinate) in the linear regime, corresponding to the initial state for our experiment, these points are disjoint and behave differently in the nonlinear regime.

## IV. DISCUSSION

There are two notable experimental results interesting for interpretation. First, the distinctive behaviour (the movement from the initial position towards the geometrical focus and back to transducer) of the on-axis pressure maximum point under the increasing driver voltage, and the correlation of this behaviour with the shock formation in the wave profile. Second, the different behaviour of the pressure and intensity maximum position in the nonlinear regime.

The physical basis of the movement of the maximum pressure point towards the geometrical focus in the nonlinear regime is as follows: there is an initial range of the driving voltage (in our case the upper level of this range being 250 Vpp) for which the waveform presents incremental nonlinear distortions without shock front formation [see Fig. 2(b)]. The last circumstance implies the absence of the nonlinear absorption effect (the linear thermoviscous absorption of water negligible and does not play an essential role here). In this case the determinant effect is the appearance of the higher harmonics, which implies an increase of the beam effective frequency, and therefore a decrease of the diffraction effect (since the effective Rayleigh length increases). This leads to the movement of the real focus toward the geometrical focus. Note that for an infinitely large frequency (geometrical acoustics limit) the real focus should coincide with geometrical focus $R$.

The further increase of the transducer driving voltage (within the range 250–500 Vpp in our case) is accompanied by the formation and development of shock fronts on the waveform [see Fig. 2(c)] that causes a great absorption (so-called high-frequency nonlinear absorption). This effect gives the opposite result: the high-frequency nonlinear absorption depresses mainly the highest harmonics and so the effective beam frequency decreases, the diffraction effect again grows and the effective (real) focus moves aside toward the transducer. This is the simplest interpretation of the mentioned effects; a more detailed analysis can be found in [6].

The cause of the second notable experimental fact (the small displacement of the intensity maximum point, and the discrepancy with the pressure maximum position) is the special character of the nonlinear deformation of time profiles shown in Figs. 2(b) and 2(c), for which the quite fast growth of the profile peak together with its fast narrowing is typical. This process corresponds to a deceleration in the increase of the area under this peak. Under the condition [3,7]

$$\int_0^T p_i(t,r,z)dt = 0 \tag{6}$$

the growth of the area of the negative part of the profile is also decelerated, and the intensity (as the square of the full area under the profile curve, see Eq.(5)) slows down in comparison with the growth of the pressure peak value. This becomes apparent in the lag of nonlinear shift of the intensity maximum compared with the shift of pressure maximum, shown in Fig. 2(a).

One can suppose the existence of specific cases (with certain initial beam structures) where the nonlinear increase of the peak pressure be accompanied by a decrease of the area under the curve profile. In this case, under the increasing of the nonlinearity degree, the on-axis intensity maximum should move towards the transducer, whilst the pressure maximum should move in the opposite direction. One example of this behaviour is found in the well-known Gaussian beam. For the study of the nonlinear behaviour of Gaussian beams we start from the analytical pressure distributions obtained in [8] for the paraxial region and for the nonlinear regime before the shock front formation. There, analytical solutions of the Khokhlov-Zabolotskaya equation were obtained in the form $P = f^{-1}\sin(\theta_0 + \eta) + \Delta$, where $P = p/p_0$ is the acoustic pressure normalized to its peak value at the source, $\theta_0 = \tau + g\sin(\theta_0 + \phi) + \delta$ is an implicit function of time, and $f$, $\eta$, $\Delta$, $g$, $\phi$, $\delta$ are functions of the linear gain $G$, the nonlinearity parameter $N$ (ratio of geometrical focal distance to plane-wave shock formation distance) and the normalized axial distance $\sigma = z/R$, defined by Eqs. (38)-(43) in Ref. 8. The peak pressure is readily obtained as

$$P(\sigma) = \frac{1}{f} + \Delta, \tag{7}$$

where

$$f = \sqrt{(1-\sigma)^2 + (\sigma/G)^2} \tag{8}$$

and

$$\Delta = \frac{N}{2f^2(1+G^2)}\left\{\arctan\left(\frac{\sigma/G}{1-\sigma}\right)\left[\sigma - G^2(1-\sigma)\right] - G\ln f\right\}. \tag{9}$$

On the other hand, the on-axis normalized intensity distribution can be obtained, according to Eq. (5), integrating the square of the implicit solution given by Eq. (7), and results

$$I(\sigma) = \frac{1}{2f^2} - \Delta^2. \tag{10}$$

Both distributions depend only on the parameters $G$ and $N$ (defining the starting beam), and the axial coordinate $\sigma$.

In Fig. 3 the curves corresponding to the distributions (7) and (10) are depicted for $G = 1$ ($N_F = \pi^{-1}$) and different values of the nonlinearity parameter, $N = 0, 0.5, 1.0$ and $1.3$. The curves in Fig. 3(a) correspond to the on-axis pressure peak distributions given by Eq. (7), while the curves in Fig. 3(b) correspond to the on-axis intensity distributions, Eq. (10). The opposite movements of the maxima (marked with black squares in the pictures for clarity) of these two characteristics under increasing values of the nonlinearity are observed. Note that the absolute decrease of the intensity (and its maximum) in Fig. 3(b) is a consequence of the normalization of these curves. This is an indicator of the redistribution of the intensity across the beam (relative decrease of intensity in the center of the beam, and an increase on the periphery during the increase of nonliner regime)

## V. CONCLUSION

Results and conclusions shown in this work are relevant in several aspects: first of all this is a new contribution to the general theory of focused beams, and also the simultaneous comparison of the behavior of the pressure and intensity maximum position is necessary to analyze the processes determined by each parameter separately or together. For example, cavitation is determined by the maximum of the pressure and, in this sense, it is very important a precise knowledge of the position of the main pressure maximum in the focused beam. On the other hand, the intensity distribution in the beam is related with the heat deposition in the medium and intensity maximum position corresponds to the local maximal heat source. It is also interesting that there are some processes that appear just with the simultaneous existence of both sources. For example, in ultrasound surgery, the heated necrotic tissue is formed under the heat dissipated with the perturbative action of the cavitation process. For these applications the results of our workcan be relevant.

**ACKNOWLEDGMENTS**

This work was supported by the Ministerio de Educación y Ciencia of the Spanish Government, under the project FIS2005-07931-C03-02. Y.N. Makov acknowledges the support from Universidad Politécnica de Valencia (Programa de Apoyo a la Investigación y Desarrollo 2007).

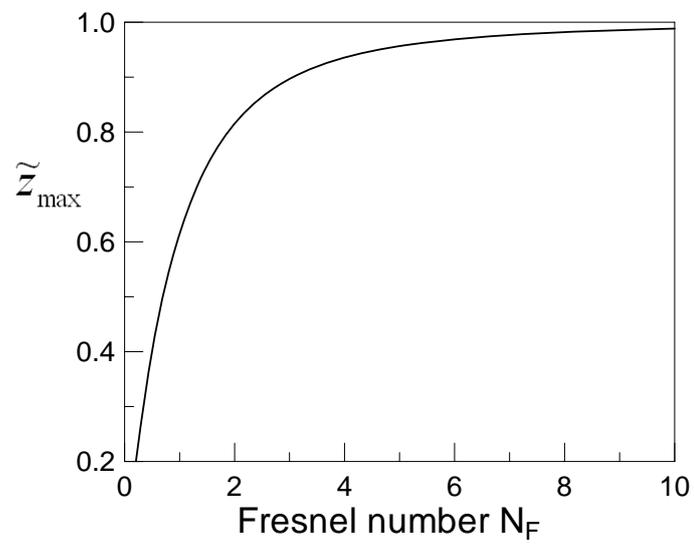

Figure 1.  Dependence of the position of the on-axis main pressure maximum $\tilde{z}_{max}$ on Fresnel number $N_F$.

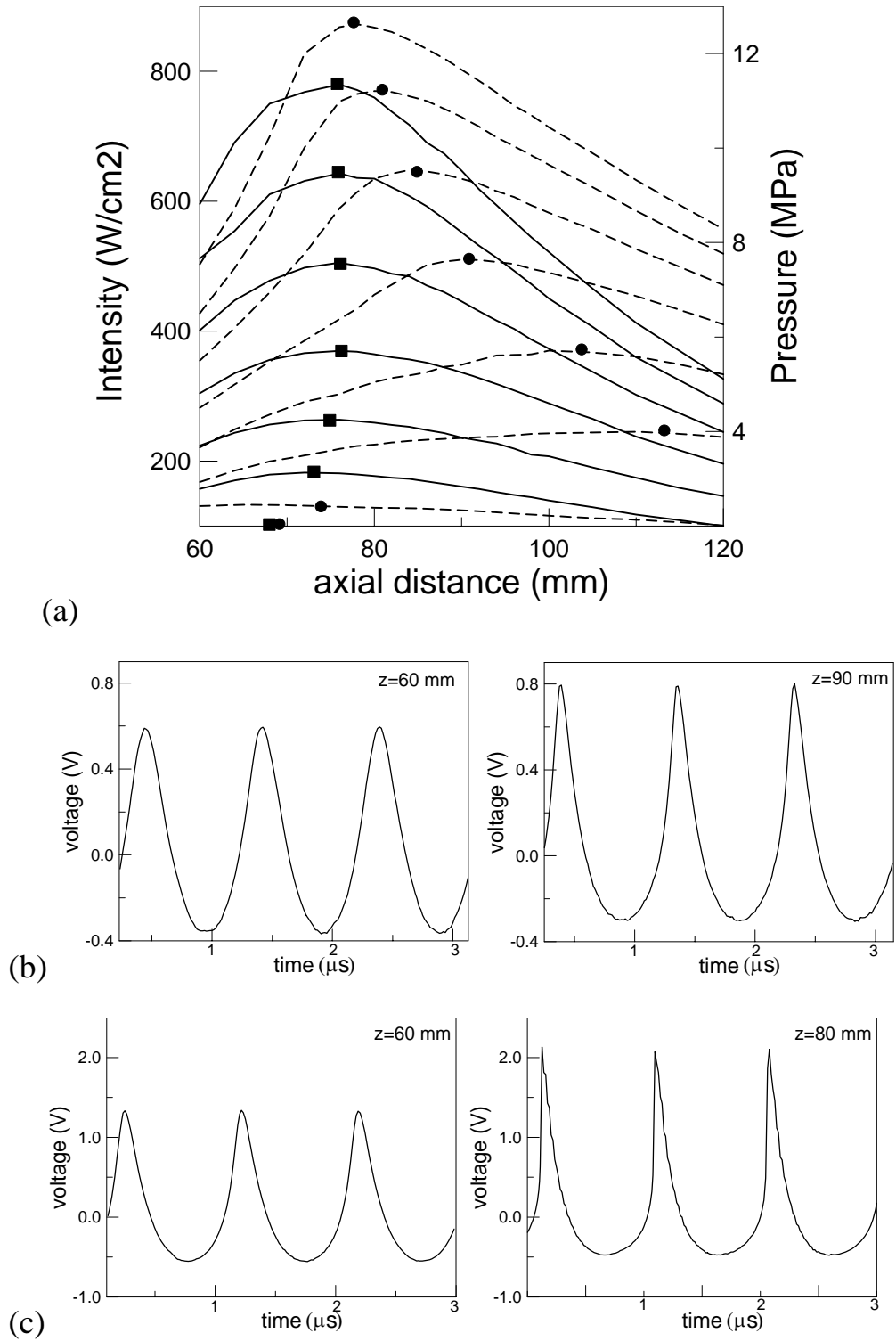

Fig. 2. (a) On-axis peak pressure (dashed lines, right axis) and intensity (continuous lines, left axis) curves. Only the neighbourhood of the maxima is plotted. Maximum values are marked with symbols. Input values are 200, 250, 300, 350, 400, 450 and 500 Vpp from bottom to top. (b) time profiles for 250 Vpp at z = 60 mm (left) and z = 90 mm (rigth). (c)

time profiles for 500 Vpp at z = 60 mm (left) and z = 80 mm (rigth). Note the waveform distortion during propagation for moderate (b) and large (c) voltage drivings.

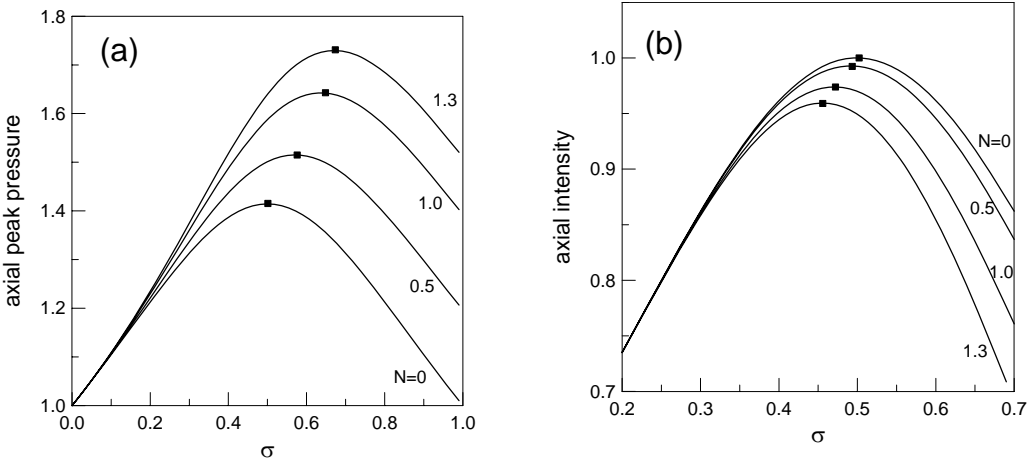

Fig. 3. On-axis peak pressure (a) and intensity (b) as follows from the analytical solutions of the KZK equation [Eqs. (7) and (10)] with an initial Gaussian distribution, for $G = 1$ and different values of the nonlinearity. The maxima are indicated with symbols. Note that the motions of pressure and intensity maxima occur in opposite directions, under the increasing of the nonlinearity.